\documentclass[aps,pra,twocolumn, superscriptaddress,amssymb,amsmath]{revtex4-1}

\usepackage{graphicx}
\usepackage{braket}

\usepackage[bookmarks=false]{hyperref}

\usepackage{float}

\begin{document}

%\preprint{}

\title{Electronic structure of negative charge transfer CaFeO\textsubscript{3} across the metal-insulator transition}


\author{Paul C. Rogge}
%\thanks{}
\affiliation{Department of Materials Science and Engineering, Drexel University, Philadelphia, Pennsylvania 19104, USA}
%\email[]{progge@drexel.edu}

\author{Ravini U. Chandrasena}
\affiliation{Department of Physics, Temple University, Philadelphia, Pennsylvania 19122, USA}
%\email[]{ravini@temple.edu}

\author{Antonio Cammarata}
\affiliation{Department of Control Engineering, Czech Technical University, Prague 166 36, Czech Republic}
%\email[]{cammaant@fel.cvut.cz}

\author{Robert J. Green}
\affiliation{Stewart Blusson Quantum Matter Institute, University of British Columbia, Vancouver, British Columbia V6T 1Z4, Canada}
\affiliation{Department of Physics \& Engineering, University of Saskatchewan, Saskatoon, Saskatchewan S7N 5E2, Canada}
%\email[]{rgreen@physics.ubc.ca}

\author{Padraic Shafer}
\affiliation{Advanced Light Source, Lawrence Berkeley National Laboratory, Berkeley, California 94720, USA}
%\email[]{PShafer@lbl.gov}

\author{Benjamin M. Lefler}
\affiliation{Department of Materials Science and Engineering, Drexel University, Philadelphia, Pennsylvania 19104, USA}
%\email[]{bml83@drexel.edu}

\author{Amanda Huon}
\affiliation{Department of Materials Science and Engineering, Drexel University, Philadelphia, Pennsylvania 19104, USA}
\affiliation{Materials Science and Technology Division, Oak Ridge National Laboratory, Oak Ridge, Tennessee 37831, USA}
%\email[]{amanda.huon@gmail.com}

\author{Arian Arab}
\affiliation{Department of Physics, Temple University, Philadelphia, Pennsylvania 19122, USA}
%\email[]{tuf04969@temple.edu}

\author{Elke Arenholz}
\affiliation{Advanced Light Source, Lawrence Berkeley National Laboratory, Berkeley, California 94720, USA}
%\email[]{earenholz@lbl.gov}

\author{Ho Nyung Lee}
\affiliation{Materials Science and Technology Division, Oak Ridge National Laboratory, Oak Ridge, Tennessee 37831, USA}
%\email[]{hnlee@ornl.gov}

\author{Tien-Lin Lee}
\affiliation{Diamond Light Source Ltd., Didcot, Oxfordshire OX11 0DE, United Kingdom}
%\email[]{tien-lin.lee@diamond.ac.uk}

\author{Slavom\'{i}r Nem\u{s}\'{a}k}
\affiliation{Peter-Gr\"{u}nberg-Institut 6, Forschungszentrum J\"{u}lich GmbH, J\"{u}lich 52425, Germany}
\affiliation{Advanced Light Source, Lawrence Berkeley National Laboratory, Berkeley, California 94720, USA}
%\email[]{s.nemsak@fz-juelich.de}

\author{James M. Rondinelli}
\affiliation{Department of Materials Science and Engineering, Northwestern University, Evanston, Illinois 60208, USA}
%\email[]{jrondinelli@northwestern.edu}

\author{Alexander X. Gray}
\affiliation{Department of Physics, Temple University, Philadelphia, Pennsylvania 19122, USA}
%\email[]{axgray@temple.edu}

\author{Steven J. May}
\affiliation{Department of Materials Science and Engineering, Drexel University, Philadelphia, Pennsylvania 19104, USA}
%\email[]{sjm95@drexel.edu}


\date{\today}

\begin{abstract}
We investigated the metal-insulator transition for epitaxial thin films of the perovskite CaFeO\textsubscript{3}, a material with a significant oxygen ligand hole contribution to its electronic structure. We find that biaxial tensile and compressive strain suppress the metal-insulator transition temperature. By combining hard X-ray photoelectron spectroscopy, soft X-ray absorption spectroscopy, and density functional calculations, we resolve the element-specific changes to the electronic structure across the metal-insulator transition. We demonstrate that the Fe electron valence undergoes no observable spectroscopic change between the metallic and insulating states, whereas the O electronic configuration undergoes significant changes. This strongly supports the bond-disproportionation model of the metal-insulator transition for CaFeO\textsubscript{3} and highlights the importance of ligand holes in its electronic structure. By sensitively measuring the ligand hole density, however, we find that it increases by ${\sim}$5-10\% in the insulating state, which we ascribe to a further localization of electron charge on the Fe sites. These results provide detailed insight into the metal-insulator transition of negative charge transfer compounds and should prove instructive for understanding metal-insulator transitions in other late transition metal compounds such as the nickelates.
\end{abstract}

%\keywords{}

\maketitle

%\section{Introduction}
%\subsection{}
%\subsubsection{}

\section{INTRODUCTION}
In order to understand and ultimately control the electronic properties of strongly correlated materials, the basic electronic structure and the relevant interactions that govern it must be known. Of particular interest is the metal-insulator transition (MIT) of correlated metal oxides, the description of which has been a long-standing challenge \cite{MIT_review, Barman_nickelate_neg_charge_transfer, Sawatzky_neg_charge_trans_1, Sawatzky_bond_disproportionation, Marianetti_site_selective_Mott}. Within metal oxides, the nickelates and ferrates are notable because ligand holes--not electrons on the transition metal--appear to be predominantly involved in the MIT. Their assignment as `negative charge transfer energy' materials highlights the energetic landscape, where it is energetically favorable for an electron to transfer from the oxygen atom to the transition metal atom, leaving an oxygen ligand hole ($\underline{L}^1$) \cite{ZSA, Sawatzky_neg_charge_trans_1}. For example, the rare-earth nickelates (\textit{RE}NiO\textsubscript{3}) nominally have seven electrons on the Ni site ($d^7$, $e_g^1$), but they are more accurately described as $d^8\underline{L}^1$ ($e_g^2$) \cite{Barman_nickelate_neg_charge_transfer, Sawatzky_neg_charge_trans_1}. In their insulating state, nickelates adopt a rock-salt super-structure of alternating dilated and contracted octahedra, and their electronic structure concomitantly transitions from $2d^8\underline{L}^1 \rightarrow d^8\underline{L}^0 + d^8\underline{L}^2$, where the ligand holes bond more strongly with the Ni in the contracted octahedra ($d^8\underline{L}^2$) leading to bond disproportionation \cite{Sawatzky_bond_disproportionation, Robert, Matsuno_CFO_dispro}. Clearly, understanding the basic electronic structure of these negative charge transfer materials is critical for describing their metal-insulator transitions as well as their overall functional responses to external fields. Moreover, because their MIT is strongly coupled to lattice distortions, determining the influence of epitaxial strain on thin films of these materials is crucial to rationally engineer non-bulk electronic properties in heterostructures of these materials. Here, we present such a picture for epitaxial films of CaFeO\textsubscript{3}, a nominal $d^4$ ($e_{g}^1$) negative charge transfer material that exhibits a MIT \cite{Kawasaki_CFO_first_transport, Bocquet_SFO_ligand_holes, Woodward_CFO, Matsuno_CFO_dispro, Takeda_CFO}.

Comparing CaFeO\textsubscript{3} (CFO) and SrFeO\textsubscript{3} (SFO) highlights the interplay between crystal structure and electronic properties typical of perovskite oxides. SFO is cubic with Fe-O-Fe bond angles of 180$^\circ$ and is metallic down to 4 K \cite{MacChesney_SFO}. Substituting Sr with the smaller isovalent Ca reduces the Fe-O-Fe bond angles to 158$^\circ$ \cite{Woodward_CFO, Takeda_CFO}, and CFO undergoes a MIT at 290 K \cite{Kawasaki_CFO_first_transport}. The MIT is accompanied by a structural phase transition from \textit{Pbnm} orthorhombic in the metallic state to $P2_1/n$ monoclinic in the insulating state \cite{Woodward_CFO, Takeda_CFO}. Within this insulating monoclinic structure, CFO adopts a rock-salt super-structure (or `breathing distortion') with alternating dilated and contracted FeO\textsubscript{6} octahedra, where the difference in Fe-O bond lengths is ${\sim}$0.1 \AA \ \cite{Woodward_CFO, Rondinelli_CFO_spin_assisted}.

Given the resemblance of CFO to the rare-earth nickelates, one may expect epitaxial strain to similarly control the metal-insulator transition temperature (T*). In nickelates, compressive strain lowers T* and can ultimately quench the MIT, and tensile strain increases T* \cite{Liu_strain_MIT_NNO, Liu_nickelate_ligand_holes_strain, Chu_NNO_MIT_strain, Triscone_SNO_MIT}. Here, however, we show that both compressive \textit{and tensile} strain lower T* for CFO, suggesting that strain affects the MIT driving force differently than in the nickelates. Combining synchrotron-based X-ray photoelectron and X-ray absorption spectroscopy, we confirm the bond-disproportionation model of the CFO MIT. We show that an energy gap opens at the Fermi level, which is consistent with our first-principles calculations. Given the importance of ligand holes in the MIT, we use X-ray absorption spectroscopy to probe the oxygen \textit{K}-edge pre-peak, which results directly from ligand holes, above and below T* and as a function of strain. We find that ligand holes are not conserved across the MIT; rather, the ligand hole density increases in the insulating state. These results, and their notable differences from the much-studied nickelates, provide important insight into strain-induced changes to negative charge transfer compounds and their metal-insulator transitions.

\section{EXPERIMENTAL}

Past synthesis of CaFeO\textsubscript{3} employed a two-step process: forming reduced CaFeO\textsubscript{3-$\delta$} followed by further annealing under high oxygen pressure (GPa) or ozone in order to achieve the relatively high oxidation state of CaFeO\textsubscript{3} \cite{Woodward_CFO, Hayashi_first_CFO_film, Akao_CFO_film, Takeda_CFO}. We employ a similar approach by depositing reduced CFO films in background oxygen and subsequently annealing the as-grown films in an oxygen plasma. 

Epitaxial CaFeO\textsubscript{3}(001)\textsubscript{pc} films of nominally 40 pseudocubic (pc) unit cells (${\sim}$15 nm thick) were grown by oxygen-assisted molecular beam epitaxy at ${\sim}$650$^\circ$C with an oxygen partial pressure of 8x$10^{-6}$ Torr. The as-grown films were subsequently annealed by heating to ${\sim}$600$^\circ$C in oxygen plasma (200 Watts, 1x$10^{-5}$ Torr chamber pressure) and then cooled in oxygen plasma by progressively turning down the heater to zero output power over approximately one hour, followed by continued exposure to the plasma for another hour to ensure complete cooling to room temperature. Compressive and tensile strain was achieved via growth on single crystal substrates: YAlO$_3$ (YAO, -2.0\% strain), SrLaAlO$_4$ (SLAO, -0.7\%), LaAlO$_3$ (LAO, 0.2\%), (La$_{0.18}$Sr$_{0.82}$)(Al$_{0.59}$Ta$_{0.41}$)O$_3$ (LSAT, 2.3\%), and SrTiO$_3$ (STO, 3.3\%). Prior to all measurements, the films were re-annealed in oxygen plasma by the same post-growth process to mitigate oxygen deficiency.

Electrical transport measurements were performed with a Quantum Design Physical Property Measurement System using a van der Pauw geometry with silver paint electrodes. Film thickness was extracted from X-ray reflectivity measurements obtained with a Rigaku SmartLab X-ray diffractometer. Reciprocal space maps and (00$l$) scans were measured with an X'Pert Pro Panalytical four-circle high resolution X-ray diffractometer.

Soft X-ray absorption spectroscopy (XAS) was performed at the Advanced Light Source beamline 4.0.2 and at the REIXS beamline at the Canadian Light Source (10ID-2). Hard X-ray photoelectron spectroscopy (HAXPES) was performed at the Diamond Light Source beamline I09 using 6.45 keV photons with an estimated probing depth of 12 nm \cite{XPS_escape_depth}.

Density functional theory calculations were performed using the projector-augmented wave (PAW) formalism \cite{Blochl_PAUW} as implemented in the Vienna Ab initio Simulation Package (VASP) \cite{Kresse_DFT_1, Kresse_DFT_2} with a minimum plane-wave cutoff of 600 eV and the revised Perdew-Burke-Ernzerhof (PBE) functional for densely packed solids \cite{Perdew_DFT_1} plus Hubbard U method (PBEsol + U) \cite{DFT_5}. We chose the spherically averaged form of the rotationally invariant effective U parameter of Dudarev et al. \cite{Sutton_DFT_1} with a $U_{eff} = 3.0$ eV \cite{Rondinelli_CFO_spin_assisted} on the Fe $d$ orbitals, which yields no significant differences in the main features of the structural phase transition for $U_{eff}$ values between 3.0 and 4.0 eV, the latter used in Ref. \cite{Satpathy_CFO_DFT}. We imposed FM order on all Fe sites and then fully relaxed the spin density. The Brillouin zone was sampled with a minimum of a 7 x 7 x 7 $k$-point mesh and integrations are performed with 20-meV Gaussian smearing. Full structural (atomic and lattice) relaxations were initiated from the neutron-diffraction data \cite{TAKANO_CFO_charge_disproportionation, Woodward_CFO} and the forces minimized to a 0.5 meV-\AA\textsuperscript{-1} tolerance.

\section{STRUCTURE AND ELECTRICAL TRANSPORT}

\begin{figure}
 \includegraphics{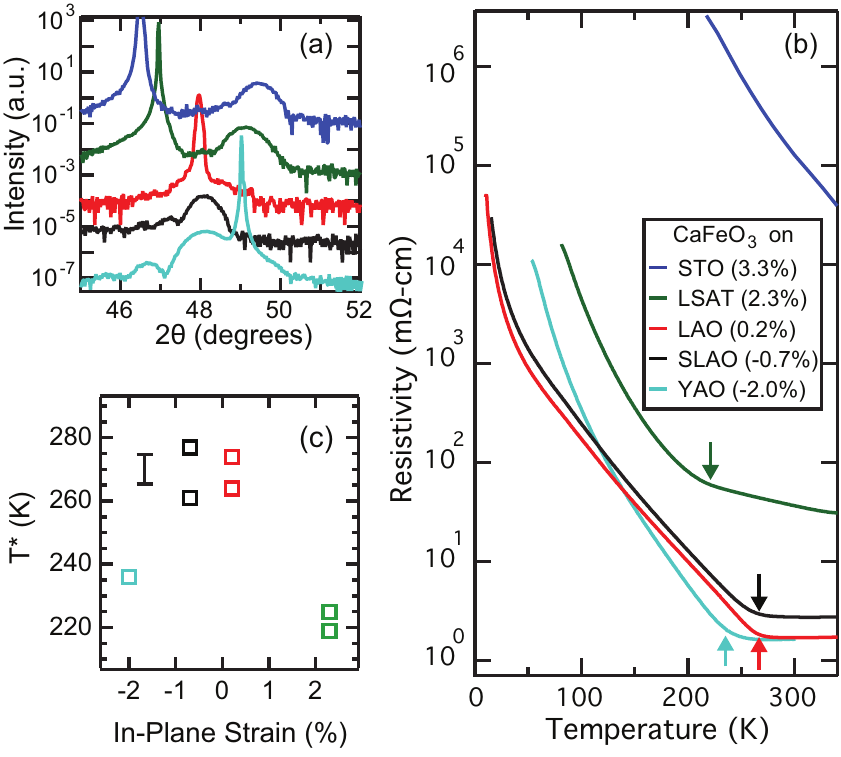}
 \caption{(a) X-ray diffraction of the CaFeO\textsubscript{3}(002)\textsubscript{pc} reflection demonstrates a strain-dependent $c$-axis parameter (see legend in (b)) for the nominally 40 pseudocubic unit cell thick films. (b) Temperature-dependent electrical resistivity of the CFO films after annealing in oxygen plasma. The arrows indicate the metal-insulator transition temperature (T*), which is plotted in (c) as a function of epitaxial strain, where the error bar is the estimated uncertainty in T*. Each data point in (c) represents a different sample.
 \label{Transport_figure}}
 \end{figure}

X-ray diffraction confirms that the films are epitaxially strained. The (002)\textsubscript{pc} film peak systematically shifts with in-plane strain, as seen in FIG. \ref{Transport_figure}(a). Reciprocal space maps of the film and substrate (103) peaks further demonstrate that the films are epitaxially strained (see Supplemental Material \cite{SI}). 

The room temperature electrical resistivity (FIG. \ref{Transport_figure}(b)) of the least strained film (CFO/LAO, 1.2 m$\Omega$-cm) is approximately equal to that of bulk CFO (3 m$\Omega$-cm) \cite{Matsuno_CFO_dispro}, indicating that the films are stoichiometric and high quality. Temperature-dependent resistivity measurements confirm that the films undergo a metal-insulator transition at T = T*, where T* is taken as the inflection point of the temperature-dependent resistivity as determined by the maximum of the second derivative \cite{SI}. T* was determined for multiple samples, and T* for each sample is shown in FIG. \ref{Transport_figure}(c).

We find that strain in CFO thin films has a large and asymmetric effect on electrical transport. Whereas compressive strain has a minor effect on the 300 K resistivity, tensile strain increases it by orders of magnitude. Moreover, in contrast to large compressive strain (CFO/YAO (-2.0\%)), large tensile strain (CFO/LSAT (2.3\%)) eliminates the strictly metallic transport above T* and significantly broadens the MIT. Surprisingly, both tensile and compressive strain lower T*. As seen in FIG. \ref{Transport_figure}(c), T* for CFO/YAO (-2.0\%) is reduced by 30 K, and T* for CFO/LSAT (2.3\%) is reduced by 45 K.  

\begin{figure}
 \includegraphics{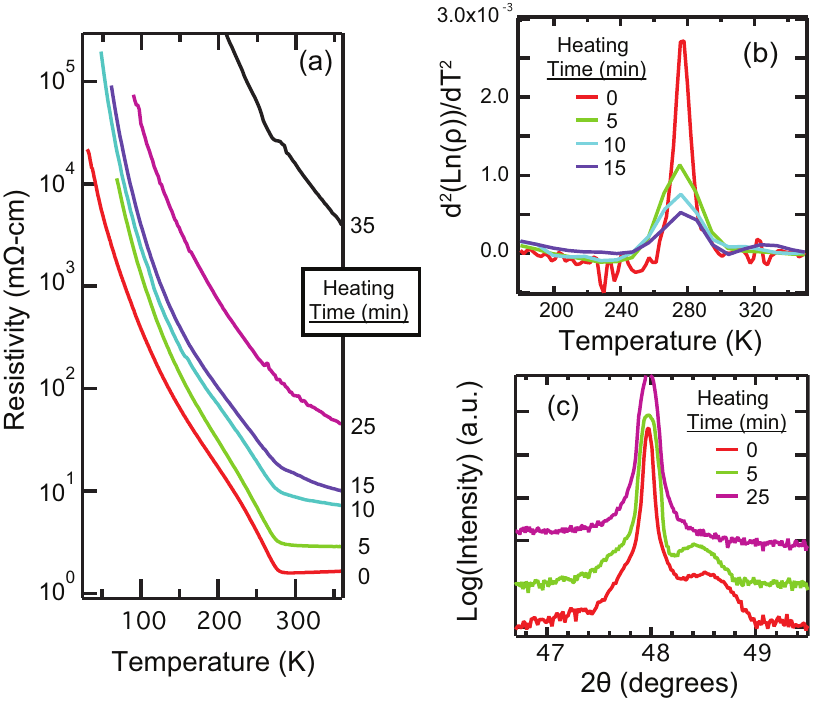}
 \caption{(a) Temperature-dependent resistivity recorded after progressively reducing a CFO/LAO film by heating to 150$^{\circ}$C for the indicated time. The second derivative of the electrical transport, shown in (b), indicates no change in T* with increased heating time. X-ray diffraction of the CFO(002)\textsubscript{pc} reflection shows the film peak moving to lower $2\theta$ with increasing heating time as the film reduces.
 \label{fig_T_star}}
 \end{figure}

We eliminate oxygen vacancies as the source of the T* suppression by evaluating T* of a nominally unstrained CFO film that is progressively reduced. A CFO/LAO film (0.2\% strain), 58 pseudocubic unit cells thick, was intentionally reduced via heating to 150$^\circ$C and characterized at various time intervals. Owing to the instability of the nominal Fe\textsuperscript{4+} valence state, ferrates are susceptible to reduction upon heating in air \cite{Xie_LSFO_reduction, SFO_oxygen_loss}. As seen in FIG. \ref{fig_T_star}(a), the initial film (heating time = 0 minutes) exhibits the expected transport behavior, and increasing heating time results in an increasing room temperature resistivity. Notably, after heating for several minutes the film still exhibits a MIT. Although the MIT broadens, T* itself remains constant as seen in FIG. \ref{fig_T_star}(b). X-ray diffraction (FIG. \ref{fig_T_star}(c)) reveals an increase in the \textit{c}-axis parameter as demonstrated by the film (002)\textsubscript{pc} peak shifting to lower 2$\theta$, which is consistent with oxygen leaving the film \cite{Adler_chemical_expansion}. Subsequently re-annealing the film in oxygen plasma recovered the original transport behavior, indicating that the film was not damaged by the heating process. Based on this relative insensitivity of T* to oxygen loss, we conclude that the T* suppression is strain-induced. 

\section{SPECTROSCOPY RESULTS}
\subsection{CaFeO\textsubscript{3} MIT: Bond Disproportionation}
 
Similar to the nickelates, the MIT of CFO has been described using different models. Early reports described the MIT in terms of charge disproportionation on the Fe sites (2Fe$^{4+} \rightarrow$ Fe$^{3+}$ + Fe$^{5+}$) \cite{TAKANO_CFO_charge_disproportionation}, whereas others noted the importance of ligand holes and proposed a MIT dictated by $2d^5\underline{L}^1 \rightarrow d^5\underline{L}^0 + d^5\underline{L}^2$ \cite{Kawasaki_CFO_first_transport, Bocquet_SFO_ligand_holes}. Given the challenge in describing the MIT of negative charge transfer perovskites, we present an extensive picture of CFO's electronic structure across its MIT. Here, we use synchrotron radiation to measure the electronic structure above and below T* to understand whether it is the electronic structure of the Fe or O sites that change the most across the MIT. By combining core level and valence band photoemission spectroscopy, X-ray absorption spectroscopy, and first-principles calculations, we develop a complete picture of CFO's electronic structure.

We first demonstrate that the spectroscopy results are consistent with a fully oxidized film. We then determine the nature of the MIT by taking advantage of the element specificity provided by synchrotron radiation to probe Fe and O separately. Of the two films with the lowest strain, we focus on CFO/SLAO (-0.7\%). However, we note that the O $K$-edge XAS data discussed below is from CFO/LAO (+0.2\%) because a full XAS spectrum from CFO/SLAO was not obtained. As will be shown later, the differences in the O $K$-edge XAS between CFO/SLAO and CFO/LAO are negligible in the context of the following discussion.

\begin{figure}
 \includegraphics{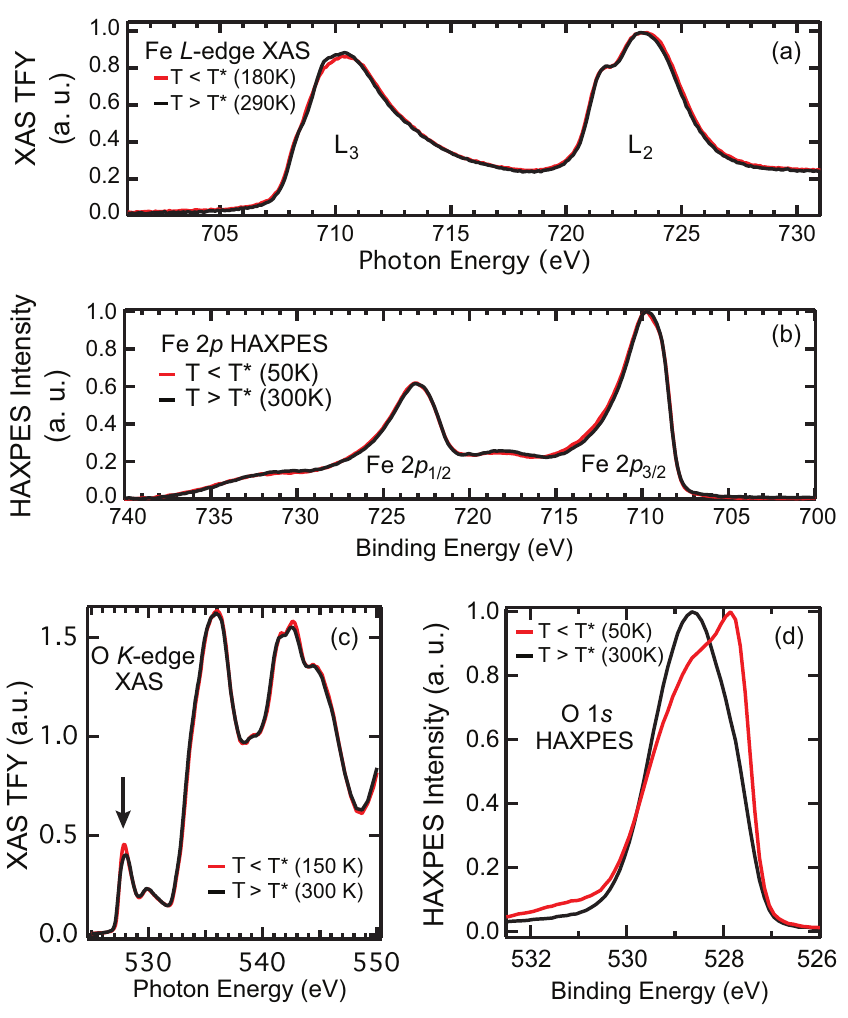}
 \caption{(a) Fe $L$-edge XAS (total fluorescence yield, TFY) and (b) Fe 2$p$ core level HAXPES measured above and below T* for CFO on SLAO(001). (c) O $K$-edge XAS (TFY) for CFO/LAO and (d) O $1s$ core level HAXPES for CFO/SLAO measured above and below T*. Note that the relative oxygen pre-peak intensity in the X-ray absorption spectra (528 eV, see arrow in panel (c)) does not accurately represent the oxidation state because the substrate contributes to the O TFY signal above the pre-peak.
 \label{XPS_XAS_figure}}
 \end{figure}

The Fe $L$-edge X-ray absorption spectra (FIG. \ref{XPS_XAS_figure}(a)) and the Fe 2$p$ core level photoelectron spectra (FIG. \ref{XPS_XAS_figure}(b)) exhibit features consistent with a predominantly ``Fe$^{4^+}$'' valence state (\textit{i.e.}, a fully oxidized film), where the quotation marks acknowledge that Fe is not strictly $d^4$ (Fe$^{4^+}$) for intrinsic CaFeO\textsubscript{3} but rather a mixture of $d^4$ and $d^5\underline{L}^1$ (strongly hybridized Fe$^{3^+}$). We are unaware of any previous core level X-ray photoelectron spectroscopy or soft X-ray absorption spectroscopy of CaFeO\textsubscript{3}, and so we compare these results to those of isoelectronic SrFeO\textsubscript{3}. Specifically, the Fe $L$-edge XAS spectra (measured by total fluorescence yield (TFY)) shows a single, broad $L_3$ peak, which is consistent with the nominal ``Fe$^{4^+}$'' valence state \cite{Abbate_SFO_XAS, Reduced_SFO_XAS}. This is in significant contrast to the Fe$^{3^+}$ XAS spectrum obtained from a EuFeO\textsubscript{3} reference sample, which exhibits two well-separated peaks for both the $L_3$ and $L_2$ edges, and is shown in the Supplemental Material \cite{SI}. Although XAS measured by TFY for transition metals exhibits strong self-absorption effects on the $L_3$ peak, resulting in its reduced intensity, TFY is bulk-sensitive whereas total electron yield (TEY) is surface-sensitive. Given the unusually high valence state of CaFeO\textsubscript{3}, it is likely that the film surface is somewhat reduced, and indeed the TEY data exhibit a small kink near 708 eV \cite{SI}. As such, we analyze the TFY data because they are more representative of the entire film. 

Additionally, the Fe $2p$ core level spectra in FIG. \ref{XPS_XAS_figure}(b) are consistent with that of SrFeO\textsubscript{3}. XPS is significantly less sensitive in distinguishing the Fe$^{3^+}$ and ``Fe$^{4^+}$'' valence states because both valence states exhibit a single peak for the $2p_{1/2}$ and $2p_{3/2}$ core levels as well as satellite features at ${\sim}$718 eV and ${\sim}$732 eV \cite{Bocquet_SFO_ligand_holes, Tsuyama_SFO_XPS, Chakraverty_BFO_XPS, Sergey_XPS_LFO}. The Fe$^{3^+}$ spectrum, however, exhibits a more pronounced satellite feature at ${\sim}$718 eV on top of a shoulder with increasing intensity from ${\sim}$715 eV to ${\sim}$720 eV. As seen in FIG. \ref{XPS_XAS_figure}(b), the CaFeO\textsubscript{3} spectra exhibit a small satellite feature at ${\sim}$718 eV with no pronounced shoulder. This is further highlighted by comparing the spectra in FIG. \ref{XPS_XAS_figure}(b) to a spectrum obtained from an Fe$^{3^+}$ reference sample (LaFeO\textsubscript{3} obtained from Ref. \cite{Sergey_XPS_LFO}) and is shown in the Supplemental Material \cite{SI}.

The O $K$-edge XAS shown in FIG. \ref{XPS_XAS_figure}(c) further supports that the films are fully oxidized. The presence of the `pre-peak' feature at 528 eV is consistent with the ``Fe$^{4+}$'' oxidation state \cite{Abbate_SFO_XAS}. Although oxygen in the substrate contributes to the TFY spectra, the substrate does not contribute to the pre-peak intensity. Importantly, the pre-peak directly probes ligand holes \cite{Abbate_SFO_XAS, Chen_cuprate_O_prepeak, Suntivich_O_Kedge_holes, Pellegrin_holes_prepeak} and its presence reflects the negative charge transfer energy of CFO.

If the MIT was completely driven by a real-space ordering of charge on the Fe sites, \textit{i.e.}, 2Fe$^{4^+} \rightarrow$ Fe$^{4^+ -\delta}$ + Fe$^{4^+ +\delta}$, one would expect significant differences in the Fe spectroscopy above and below T*. However, comparing the Fe HAXPES and Fe XAS spectra measured above and below T*, one sees that they are nearly indistinguishable. In contrast, the oxygen spectra exhibit significant differences between the metallic and insulating states. The O $1s$ core level shown in FIG. \ref{XPS_XAS_figure}(d) transitions from a single broad feature above T* into two overlapping but distinct features below T*. Moreover, the changes in the O $K$-edge XAS spectra (FIG. \ref{XPS_XAS_figure}(c)) are localized to the oxygen pre-peak region at 528 eV. 

These results support the bond-disproportionation model wherein the dominant change to the electronic structure is $2d^5\underline{L}^1 \rightarrow d^5\underline{L}^0 + d^5\underline{L}^2$ across the CFO metal-insulator transition. The O $1s$ core level energy depends on the oxygen ion's valence electronic shell. The transition from a single, broad peak to a clear doublet feature below T* is consistent with the oxygen electronic configuration undergoing a significant change \cite{Pawlak_O1s_core_level}, and previous theoretical results predict such a doublet feature when two ligand holes bond strongly with one octahedron in negative charge transfer perovskites \cite{Robert_XPS}. Additionally, the O $K$-edge XAS spectra demonstrate that ligand holes play an important role in the MIT. These spectral changes are not due to the substrate because it does not undergo any structural or electronic transition at these temperatures. Given that the Fe valence does not exhibit detectable changes across T*, these results strongly support the bond-disproportionation model of the MIT in which oxygen ligand holes order below T*, resulting in the rock-salt super-structure of alternating expanded $\left(d^5\underline{L}^0\right)$ and contracted $\left(d^5\underline{L}^2\right)$ octahedra. Such a model has also been described in terms of alternating degrees of ionicity and covalency across the Fe-O bonds \cite{Rondinelli_CFO_spin_assisted, Pickett_bond_dispro}.

 \subsection{MIT-induced changes at the Fermi level}
 
\begin{figure}
 \includegraphics{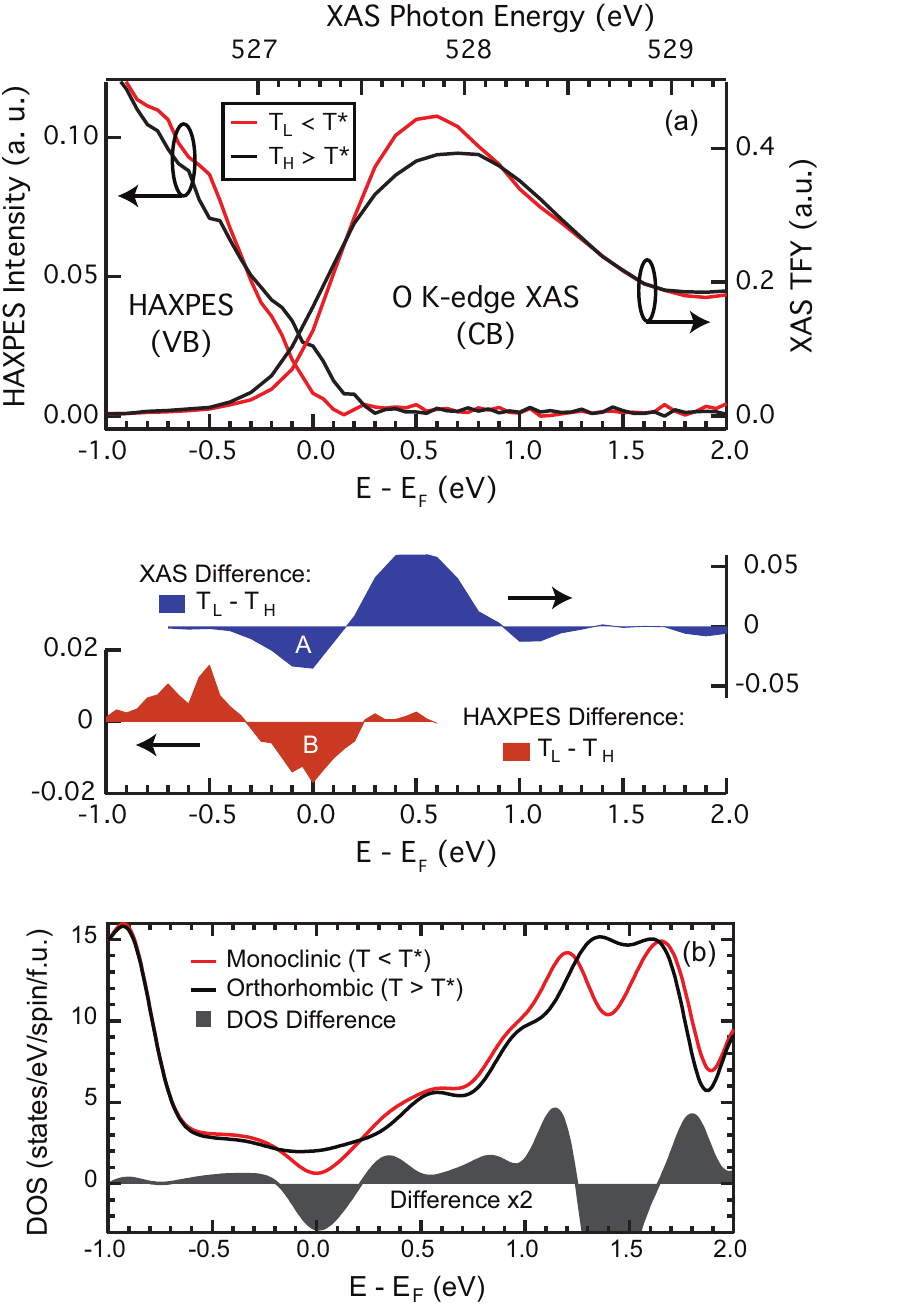}
 \caption{(a) Valence band HAXPES of CFO/SLAO measured at $T_L = 50$ K and $T_H = 300$ K, and O $K$-edge XAS of CFO/LAO measured at $T_L = 180$ K and $T_H = 290$ K. The respective difference spectra were computed by subtracting the spectrum obtained at $T_H$ from that obtained at $T_L$. In order to plot on a common axis with the HAXPES data, the O $K$-edge spectra were shifted in energy by setting the minimum of their difference spectrum to 0 eV, and the absolute energy scale for the O $K$-edge XAS is reproduced at the top for reference. Both the XAS and HAXPES spectra show a loss of states at $E_F$ (features ``A'' and ``B'' in the difference spectra, respectively) due to the opening of a band gap below T*. (b) Total DOS obtained by DFT for the monoclinic (insulating) and orthorhombic (metallic) CFO crystal structures. The DOS difference spectrum similarly shows a loss of states at $E_F$ and an increase in states just beyond the energy gap.
 \label{XPS_XAS_DFT_figure}}
 \end{figure}

Given that the electronic structure near the Fermi level, $E_F$, typically controls the relevant electronic properties, we determined the MIT-induced changes to the valence band (VB) and conduction band (CB) density of states (DOS) near $E_F$. The VB DOS was probed by VB HAXPES. Additionally, because a 1$s$ core-hole has no angular momentum and thus results in little to no Coulomb interactions with the photo-excited electron, the O $K$-edge XAS spectrum is a close approximation of the O $p$-projected density of unoccupied states, which, due to the strong Fe-O hybridization in CFO, is a good representation of the CB DOS \cite{deGroot_XAS, Suntivich_O_Kedge_holes}. Hence in FIG. \ref{XPS_XAS_DFT_figure}(a), we plot the VB HAXPES and the O $K$-edge XAS spectra near $E_F$. In order to show the change in the VB and CB across T*, the spectrum obtained above T* was subtracted from that obtained below T* (herein referred to as a `difference spectrum'). Both the HAXPES and XAS measurements were repeated to confirm reproducibility, and the binding energy was calibrated using a high-resolution Au Fermi edge measurement immediately prior to data collection.

The HAXPES difference spectrum reveals a loss of states (demonstrated by the negative intensity) at $E_F$ as CFO transitions from the metallic to the insulating phase. The positive intensity below -0.4 eV suggests that the states lost at $E_F$ have shifted to higher binding energy in the VB. This loss of states at $E_F$ agrees well with previous ultraviolet photoemission spectroscopy on bulk, polycrystalline CFO \cite{Matsuno_CFO_dispro}, but here we are able to capture the spectral changes to higher binding energies and reveal that the states at $E_F$ shift to lower energy ($E - E_F$ = -0.4 to -1.0 eV). Similarly, the O $K$-edge shifts to higher energy below T*, yielding a loss of states at $E_F$. These changes are consistent with a band gap on the order of a few hundred meV opening in the insulating state, which is in agreement with previous optical conductivity measurements of bulk CFO that showed a gap of 0.25 eV \cite{CFO_optical_gap}. Here, the combined HAXPES and XAS spectra demonstrate that both the VB and CB edges shift in energy as the band gap opens. Additionally, a comparison of the difference spectra shows that the VB edge shifts slightly more in energy than the CB edge: As seen in FIG. \ref{XPS_XAS_DFT_figure}(a), the HAXPES negative intensity at $E_F$ (labeled ``B'') has a larger full-width at half-max as compared to that exhibited by the XAS (labeled ``A''). 

These experimental observations are supported by DFT calculations of CFO's atomic and electronic structures. The relaxed structures for the monoclinic (insulating) and orthorhombic (metallic) phases were obtained as described above, and the full structural results are contained in the Supplemental Material \cite{SI}. We find that the orthorhombic structure has a single Fe site with uniform Fe-O bond lengths (${\sim}1.91$ \AA), whereas the monoclinic structure has two distinct Fe sites that exhibit a dilated (${\sim}1.97$ \AA) or contracted (${\sim}1.87$ \AA) Fe--O bond length. These results are consistent with the structural changes that accompany the bond disproportionation across the MIT, where the insulating phase consists of alternating dilated and contracted FeO\textsubscript{6} octahedra. 

The total DOS was extracted from the DFT-obtained orthorhombic and monoclinic structures, and are shown in FIG. \ref{XPS_XAS_DFT_figure}(b). To more accurately compare to the spectroscopy data, the DFT DOS were smoothed with a 0.20 eV Gaussian (FWHM) to account for the effects of total experimental resolution \cite{Rondinelli_DOS_smoothing}. The orthorhombic structure shows a non-zero DOS at $E_F$, consistent with the metallic state above T*. The monoclinic structure, however, exhibits an energy gap at $E_F$ (we note that the unsmoothed DOS fully goes to zero and shows a gap of 0.18 eV \cite{Rondinelli_CFO_spin_assisted}). Comparing to the HAXPES and XAS difference spectra, the DFT difference spectrum exhibits excellent agreement, capturing the loss of states at $E_F$ and the corresponding increase in states just beyond the gap in both the VB and CB. Good agreement is also observed when comparing the HAXPES results to the matrix-element-weighted DOS that accounts for the photoelectric cross-sections of specific orbitals \cite{SI}. Together, these experimental and first-principles results demonstrate that a band gap on the order of a few hundred meV opens at the Fermi level when CFO becomes insulating.

A closer evaluation of the O $K$-edge XAS difference spectrum reveals a notable finding: The total pre-peak intensity increases in the insulating state. The integrated intensity of the difference spectrum below 529 eV is the net change in the pre-peak intensity between the metallic and insulating states. When summing the negative and positive intensity in the difference spectrum up to 529 eV, the net change is positive. We stress that the difference spectrum accurately captures edge shifts and intensity changes, and thus the observed increased integrated intensity is not merely a repositioning of states to higher energy due to the edge shift (band gap opening) but rather is due to an increase in the total pre-peak intensity. Because the pre-peak intensity is proportional to the ligand hole density \cite{Abbate_SFO_XAS, Chen_cuprate_O_prepeak, Suntivich_O_Kedge_holes, Pellegrin_holes_prepeak}, the change in its intensity captures the change in ligand hole density. This increase in the pre-peak intensity implies that the ligand hole density is not conserved across the MIT.
 
\begin{figure}
 \includegraphics{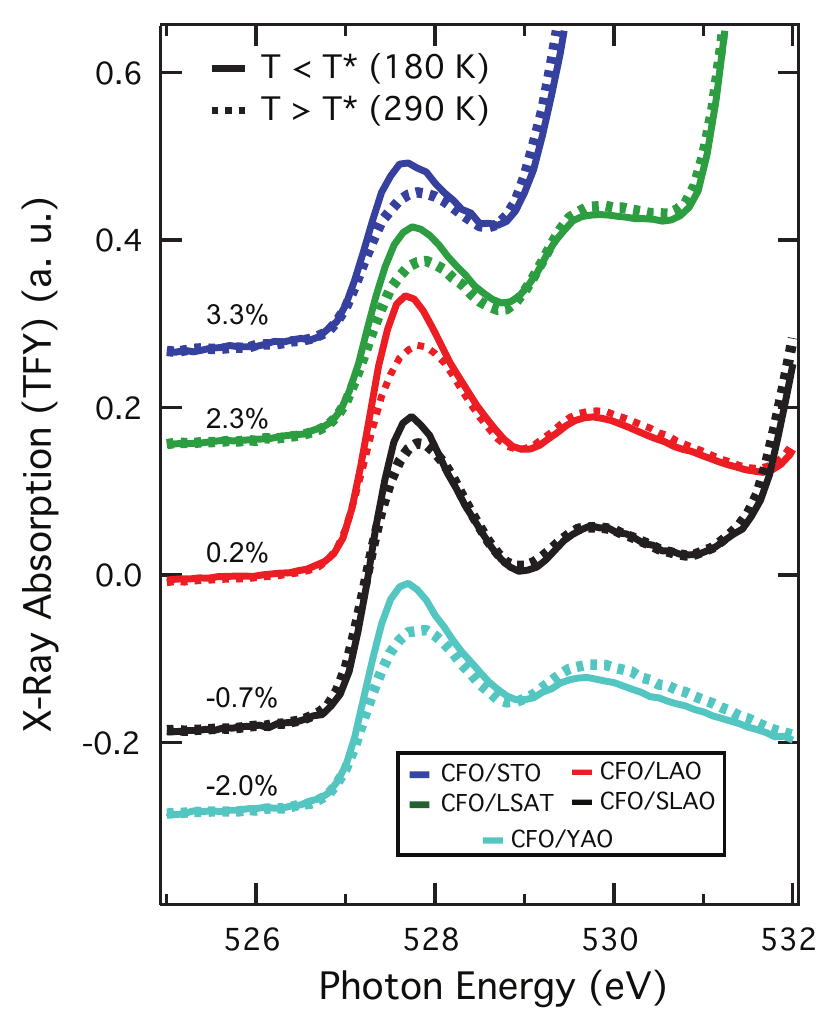}
 \caption{Oxygen $K$-edge pre-peak as measured by X-ray absorption (total fluorescence yield) above T* (dashed lines) and below T* (solid lines) for the various strained CFO films. Note that due to substrate contributions to XAS measured by TFY, absolute intensities of the pre-peak cannot be compared across different films.  \label{fig_XAS_prepeak}}
 \end{figure}

Notably, all strained films exhibit a similar increase in the oxygen ligand hole density in their insulating state. As seen in FIG. \ref{fig_XAS_prepeak}, the O $K$-edge pre-peak exhibits a higher intensity in the insulating state. (Although we could not identify a phase transition in CFO/STO from electrical transport due to its high resistivity, we include the results here for completeness.) Comparing the pre-peak areas above and below T*, the increase in the pre-peak intensity is ${\sim}$5--10\%. For this comparison we focus only on the intensity of the pre-peak feature, which is emphasized by aligning the spectra in FIG. \ref{fig_XAS_prepeak} so that this feature appears at the same energy for all samples.

We ascribe this increase in the O $K$-edge pre-peak intensity to electron transfer from the oxygen ligands to the Fe sites in the insulating state. Due to the strong Fe-O hybridization, charge is shared between Fe and O. If the hole density increases on the oxygen sites, as demonstrated by the increased pre-peak intensity, then the complementary electron density transfers to the Fe sites. Although the electronic structure is a superposition of all $\ket{d^m \underline{L}^n}$ configurations \cite{Robert}, from the very simple view of the electronic structure transforming from $\alpha \ket{d^4} + \beta \ket{d^5 \underline{L}^1}$ in the metallic state to $\alpha \ket{d^4} + \beta \left(\frac{1}{\sqrt{2}} \ket{d^5 \underline{L}^0} + \frac{1}{\sqrt{2}}\ket{d^5 \underline{L}^2}\right)$ in the insulating state, our observation implies a scenario where $\beta$ ($\alpha$) increases (decreases) upon cooling into the insulating state. In other words, the charge transfer energy becomes more negative as CFO enters the insulating state.

A rough estimate of the maximum amount of electron transfer to Fe in the insulating state is provided by examining the limit of the metallic phase being purely $\ket{d^5 \underline{L}^1}$. In such a limit, the pre-peak represents a single ligand hole per FeO\textsubscript{6} octahedron, and the observed ${\sim}$5--10\% increase in the pre-peak intensity would correspond to a transfer of 0.05--0.10 electrons to all Fe sites. Although one may expect changes in the Fe $L$-edge XAS due to this charge transfer, we note that the O $K$-edge pre-peak is a direct probe of the ligand hole density, whereas the Fe $L$-edge probes the total Fe valence and is thus less sensitive to this small change in charge density. Rather, the Fe $K$-edge is expected to better reflect the change in ligand hole density across the MIT because it, like the O $K$-edge, has an isolated pre-peak feature that is sensitive to ligand holes \cite{SFO_Fe_K_edge_XAS} and can be confirmed in future work.

\section{DISCUSSION}
The CFO MIT appears to be rather robust to epitaxial strain. Although ${\sim}2\%$ strain suppresses T* by 40 K, all films (other than CFO/STO) exhibit a phase transition in their electrical transport. Moreover, all of the strained films exhibit a similar increase in the O $K$-edge pre-peak intensity in the insulating state. This robustness is in marked contrast to some rare-earth nickelates. For example, minor compressive strain suppresses T* by over 100 K in NdNiO\textsubscript{3} (-0.3\% strain) and over 200 K in SmNiO\textsubscript{3} (-0.9\% strain) \cite{Liu_strain_MIT_NNO, Chu_NNO_MIT_strain, Triscone_SNO_MIT}. Such behavior has been explained by examining the Ni--O hybridization: Compressive strain increases Ni--O hybridization and resultantly lowers T*, and vice versa for tensile strain \cite{Rondinelli_Spaldin_Adv_Mater, Liu_nickelate_ligand_holes_strain}. 

This simple hybridization picture apparently does not directly translate to CFO, where both compressive and tensile strain lower T*. Although it is possible that both tensile and compressive strain act to increase hybridization, it seems unlikely given that CFO, like the nickelates, is a typical orthorhombic perovskite above T*. Furthermore, compared to the compressively-strained CFO film (CFO/YAO, -2.0\%), the tensile-strained CFO film (CFO/LSAT, 2.3\%) exhibits a broader MIT and an order of magnitude higher room temperature resistivity, which suggests that tensile and compressive strain are indeed acting differently. Similarly, the relative insensitivity of T* to epitaxial strain is another surprising departure from the nickelates. We surmise that the simple picture of strain-induced modification of Fe--O hybridization is further complicated by the significantly more negative charge transfer energy of the ferrates compared to the nickelates. These results, though, highlight the challenge in determining the important interactions that control the electronic structure in these strongly hybridized systems and motivate future efforts to uncover them.

\section{CONCLUSION}
We have combined hard and soft X-ray synchrotron radiation to probe the metal-insulator transition of CaFeO\textsubscript{3} as a function of epitaxial strain. The results strongly support the bond-disproportionation model of $2d^5\underline{L}^1 \rightarrow d^5\underline{L}^0 + d^5\underline{L}^2$. The opening of a band gap at the Fermi level is observed for the insulating state and supported by density functional theory calculations. By probing the oxygen ligand hole density via X-ray absorption spectroscopy, the insulating state is shown to support a ${\sim}$5-10\% higher density of ligand holes, which we attribute to a small amount of electron transfer to the Fe sites estimated to be no more than ${\sim}0.10$ electrons per Fe. Although epitaxial strain lowers the metal-insulator transition temperature, it is significantly less sensitive to strain compared to the negative charge transfer rare-earth nickelates. These results provide further insight into the role of ligand holes in the metal-insulator transition of negative charge transfer materials and are important for guiding future efforts to produce ferrate heterostructures. 

\begin{acknowledgments}
PCR, AH, and SJM were supported by the Army Research Office, grant number W911NF-15-1-0133; AXG, RUC, and AA acknowledge support from the U.S. Army Research Office, under Grant No. W911NF-15-1-018; JMR was supported by the National Science Foundation through DMR-1729303. Film synthesis at Drexel utilized deposition instrumentation acquired through an Army Research Office DURIP grant (W911NF-14-1-0493). This work used resources at the Advanced Light Source, which is a DOE Office of Science User Facility under contract no. DE-AC02-05CH11231, and at the Canadian Light Source, which is funded by the Canada Foundation for Innovation, NSERC, the National Research Council of Canada, the Canadian Institutes of Health Research, the Government of Saskatchewan, Western Economic Diversification Canada, and the University of Saskatchewan. We thank Diamond Light Source for access to beamline I09 (proposal number SI17824) that contributed to the results presented here. X-ray diffraction performed at ORNL was supported by the U.S. Department of Energy, Office of Science, Basic Energy Sciences, Materials Sciences and Engineering Division.
\end{acknowledgments}

%\bibliography{Oxide_bibliography_bibtex}


% ****** SUPPLEMENTAL  *********

\pagebreak
\onecolumngrid
\begin{center}
\textbf{\large Supplemental Material: The electronic structure of negative charge transfer CaFeO\textsubscript{3} across the metal-insulator transition}
\end{center}

\setcounter{equation}{0}
\setcounter{figure}{0}
\setcounter{table}{0}
\setcounter{page}{1}
\makeatletter
\renewcommand{\theequation}{S\arabic{equation}}
\renewcommand{\thefigure}{S\arabic{figure}}
\renewcommand{\thetable}{S\arabic{figure}}
\renewcommand{\bibnumfmt}[1]{[S#1]}
\renewcommand{\citenumfont}[1]{S#1}

\begin{figure}[H]
\includegraphics{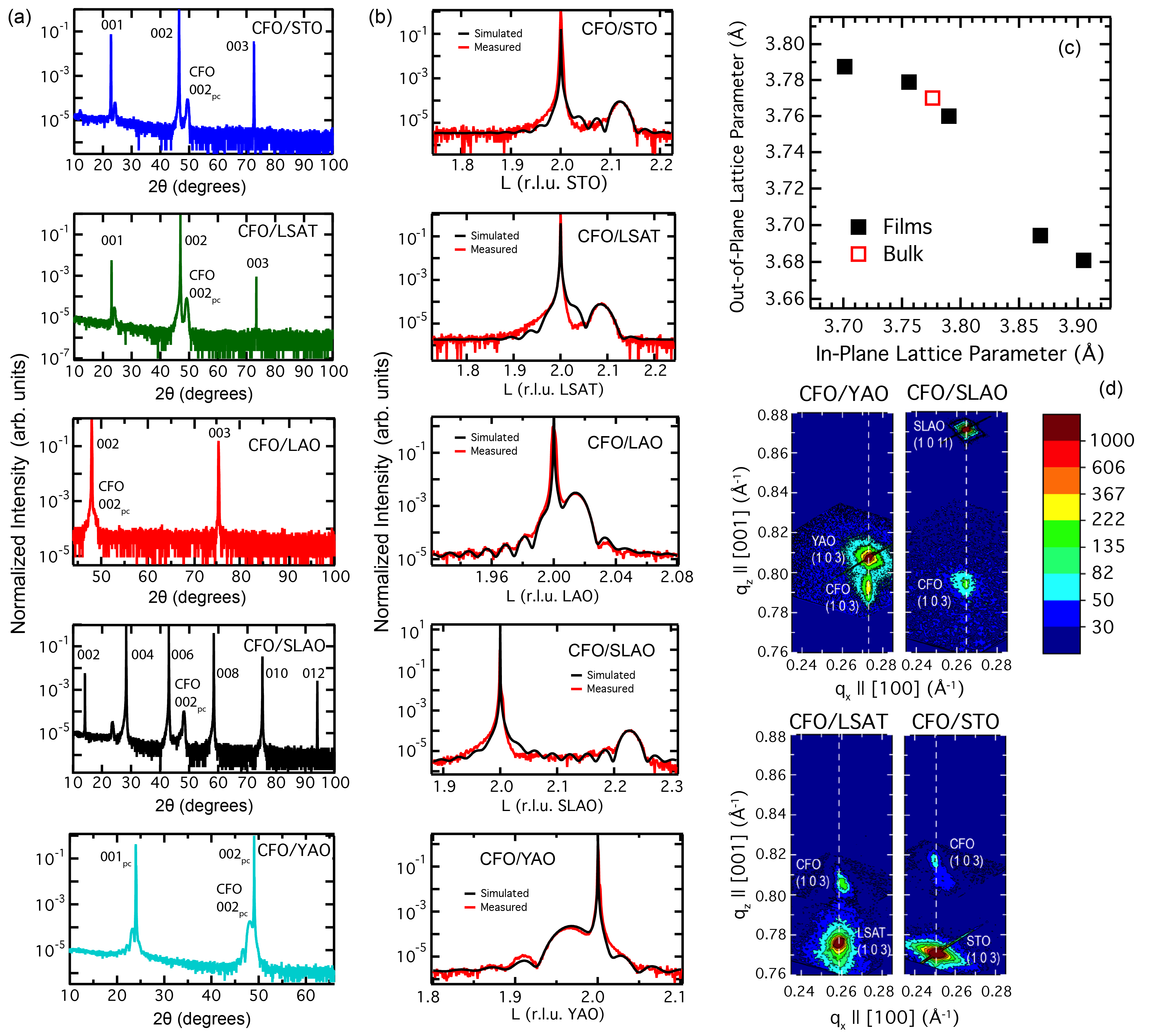}
\caption{\textit{X-Ray Diffraction Results} -- (a) $(00l)$ scans across the full $2\theta$ range confirm that the films are single phase for the 40 pseudocubic (pc) unit cell thick films. The substrate and CFO (002)\textsubscript{pc} reflections are labeled. (b) Pseudocubic $(002)$ peaks plotted along with a simulated diffraction pattern generated by GenX, showing excellent agreement. The CFO $c$-axis parameter was extracted from the simulated diffraction patterns and are plotted in (c). The CFO/LAO $c$-axis parameter was extracted by measuring a thicker (86 pseudocubic unit cells) film in order to achieve a more prominent film peak and thus a more robust fitting. (d) Reciprocal space maps of the substrate (103) and the CFO (103)\textsubscript{pc} reflections confirm that the films are epitaxially strained. 
\label{Fig_XRD_SM}}
\end{figure}

 \begin{figure}[H]
  \centering
 \includegraphics{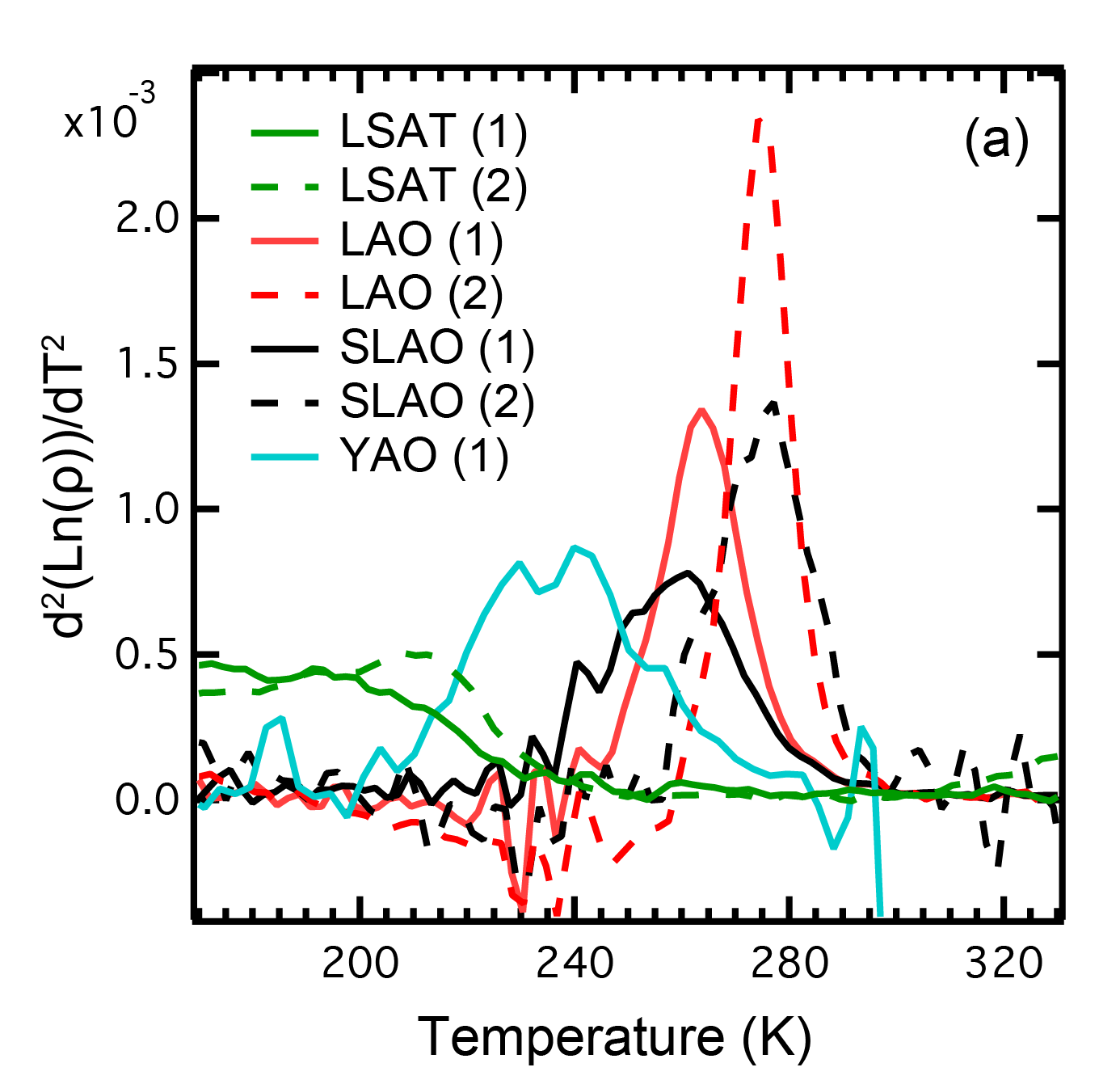}
 \caption{\textit{Metal-insulator transition temperature (T*) determination} -- The second derivative of the natural log resistivity was taken with respect to temperature and is plotted versus temperature. The temperature at which the peak maximum occurs is taken as T*. Because the CFO/LSAT resistivity does not exhibit a discrete peak in its second derivative but rather a step-like increase, the temperature at the midpoint of this step-like increase was taken as T*. The uncertainty in T* is estimated to be $\pm 5$ K. Results from multiple samples are shown.
 \label{Fig_Tstar_SM}}
 \end{figure}
 
  \begin{figure}[H]
  \centering
 \includegraphics{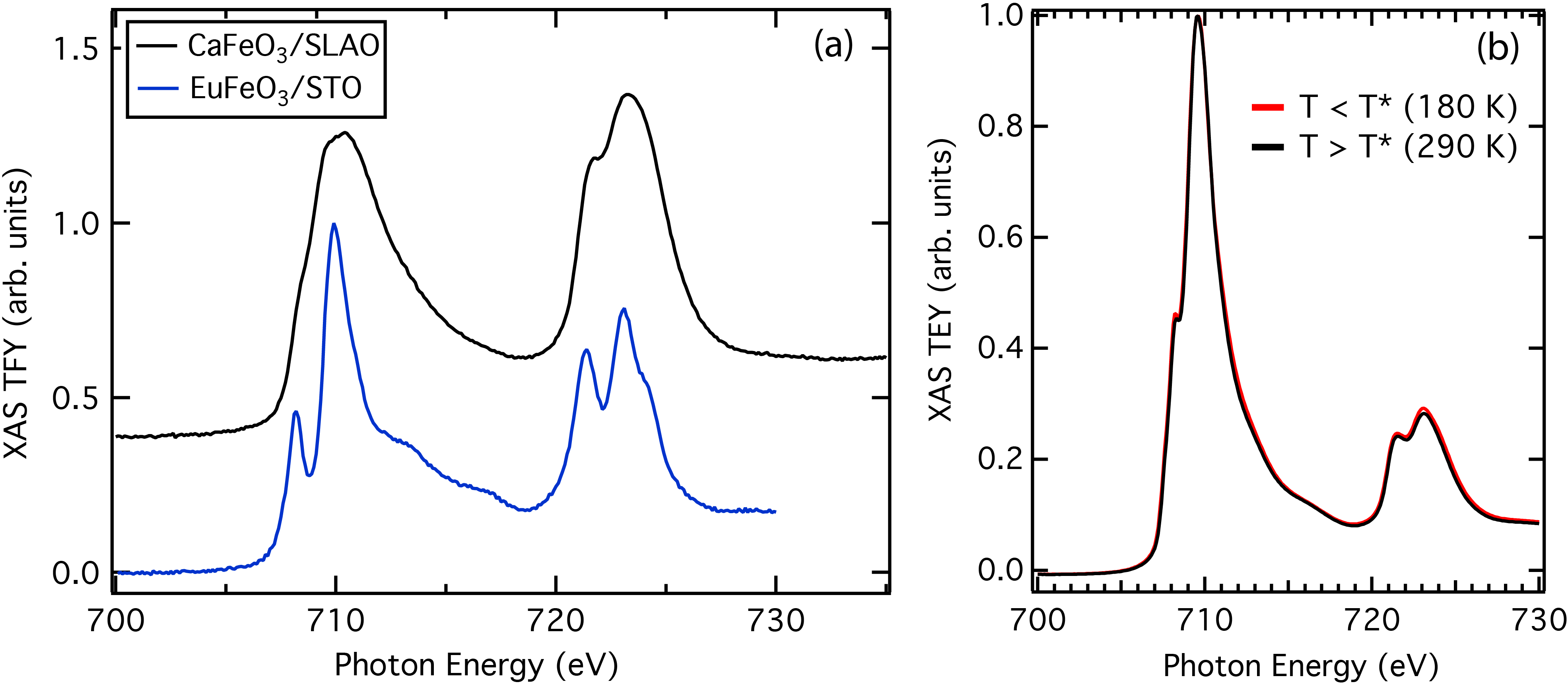}
 \caption{(a) Comparison of the Fe $L$-edge X-ray absorption for CaFeO\textsubscript{3} and an Fe\textsuperscript{3+} reference sample, EuFeO\textsubscript{3}, measured by total fluorescence yield (TFY). The single, broad $L_3$ peak for CaFeO\textsubscript{3}, compared to the clear doublet feature for Fe\textsuperscript{3+} in EuFeO\textsubscript{3}, is consistent with a fully oxidized, ``Fe\textsuperscript{4+}'', CaFeO\textsubscript{3} film. (b) X-ray absorption spectra across the Fe $L$-edge for CFO/SLAO measured by total electron yield (TEY). The small kink at ${\sim}708$ eV is suggestive of a small Fe$^{3+}$ contribution to the spectra due to oxygen loss from the film surface. No significant changes are seen between the low temperature (insulating) and high temperature (metallic) spectra. 
 \label{Fig_XAS_TEY}}
 \end{figure}
 
 \begin{figure}[H]
  \centering
 \includegraphics{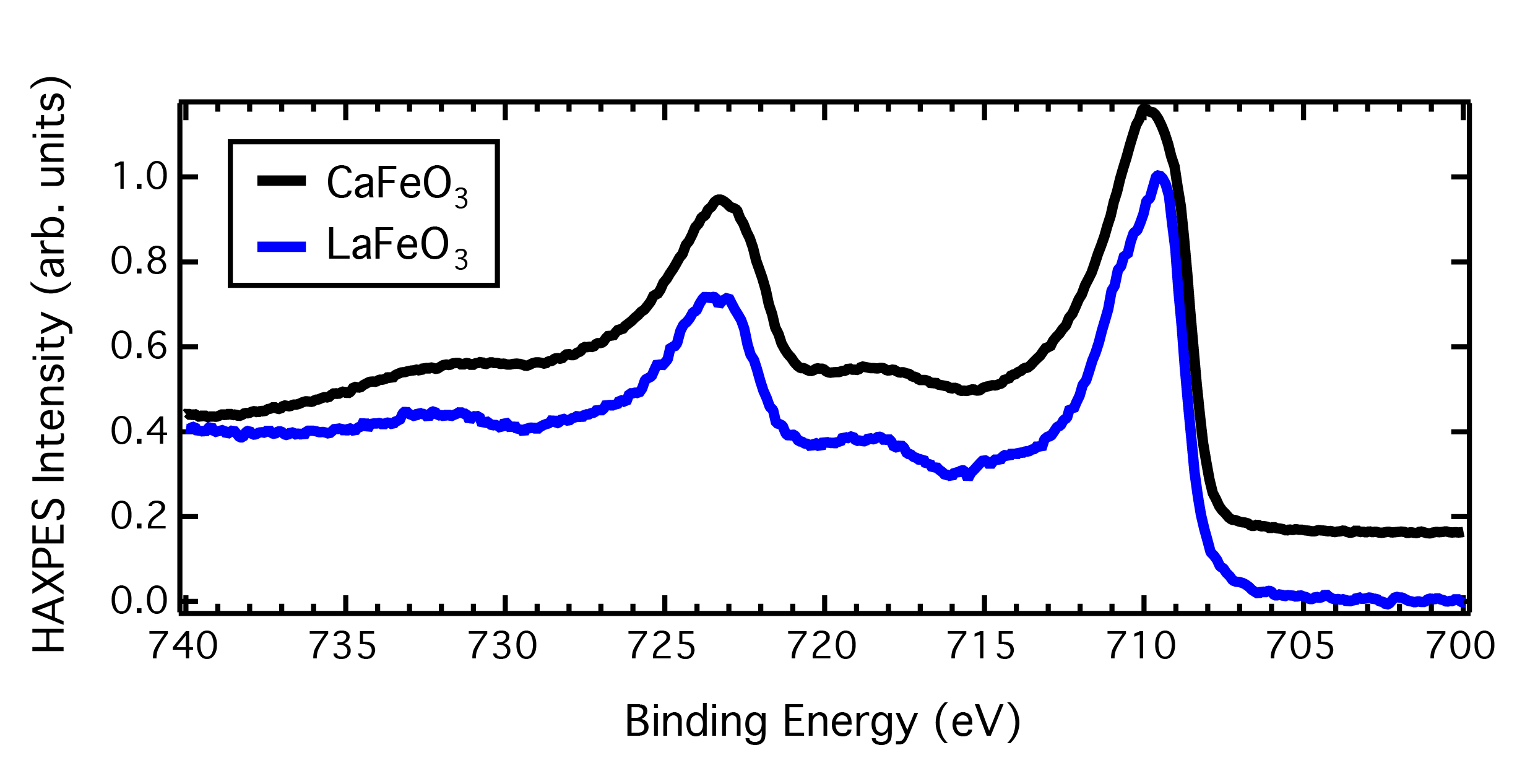}
 \caption{Comparison of the CaFeO\textsubscript{3}/SLAO Fe $2p$ HAXPES spectrum to an Fe\textsuperscript{3+} spectrum measured for a reference sample (LaFeO\textsubscript{3} film on STO obtained from Smolin et al., Adv. Mater. Interf. 4, 1700183 (2017)). Compared to the CaFeO\textsubscript{3} spectrum, the Fe\textsuperscript{3+} of LaFeO\textsubscript{3} exhibits more pronounced satellite peaks at ${\sim}$718 eV and ${\sim}$732 eV as well as a larger shoulder due to an increasing intensity from ${\sim}$715 eV to ${\sim}$720 eV.
 \label{Fig_XPS_comparison}}
 \end{figure}
 
 % \pagebreak
  
\begin{figure}[H]
 \centering
 \includegraphics{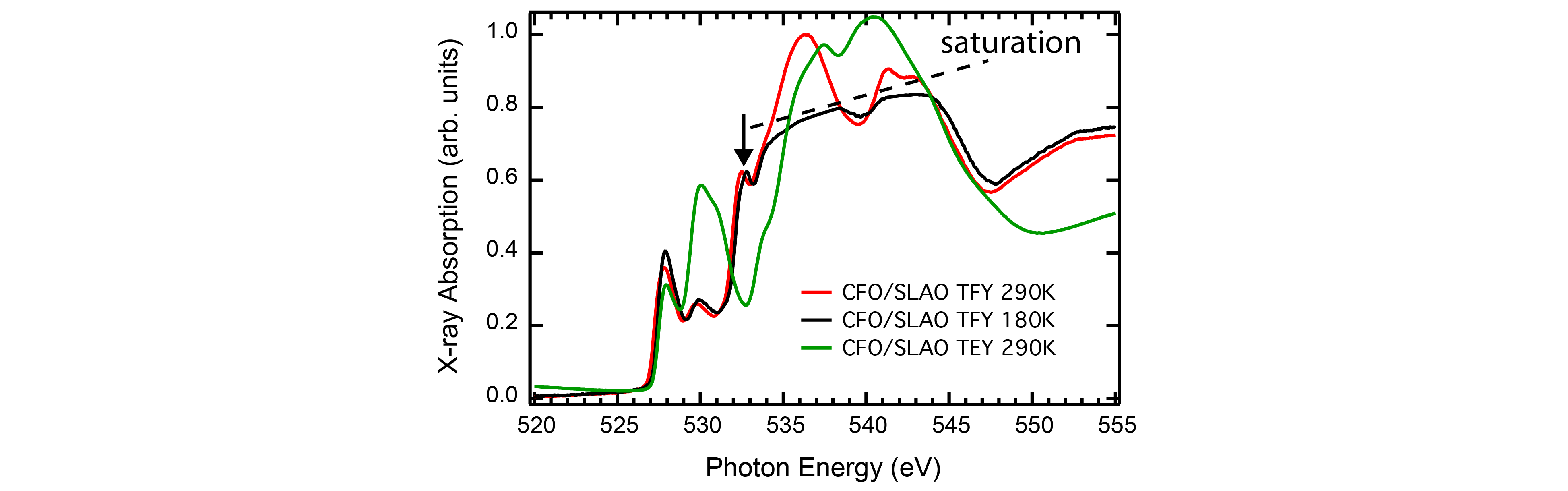}
 \caption{X-ray absorption spectra for CFO/SLAO. The 180 K TFY data were normalized by setting the intensity of the peak at 532 eV (indicated by the arrow) equal to the intensity of the same peak in the 290 K TFY spectrum.
 \label{Fig_SLAO_XAS_SM}}
 \end{figure}

\textit{X-Ray Absorption Normalization} -- The X-ray absorption data were normalized by the incident intensity as measured by an upstream gold mesh. The low energy baseline was set to unity and then zero, followed by normalizing the maximum intensity to unity. Spectra were taken in both horizontal and vertical polarizations, which were then averaged together. The CFO/SLAO O $K$-edge spectrum taken at 180 K was normalized differently due to saturation of the detector. FIG. \ref{Fig_SLAO_XAS_SM} shows the saturation (denoted by the dashed line) occurring above 534 eV. By comparing to the total electron yield (TEY) spectrum, which measures absorption from the film only, one sees a peak at $\sim$532 eV in the total fluorescence yield (TFY) spectrum, which probes the film and substrate. The presence of this peak in the TFY spectra but not the TEY spectrum indicates that it is from the SLAO substrate, which is not expected to change with temperature. Thus, the 180 K CFO/SLAO TFY spectrum was normalized such that this peak was set equal to that in the CFO/SLAO 290 K spectrum. 

  \begin{figure}[H]
   \centering
 \includegraphics{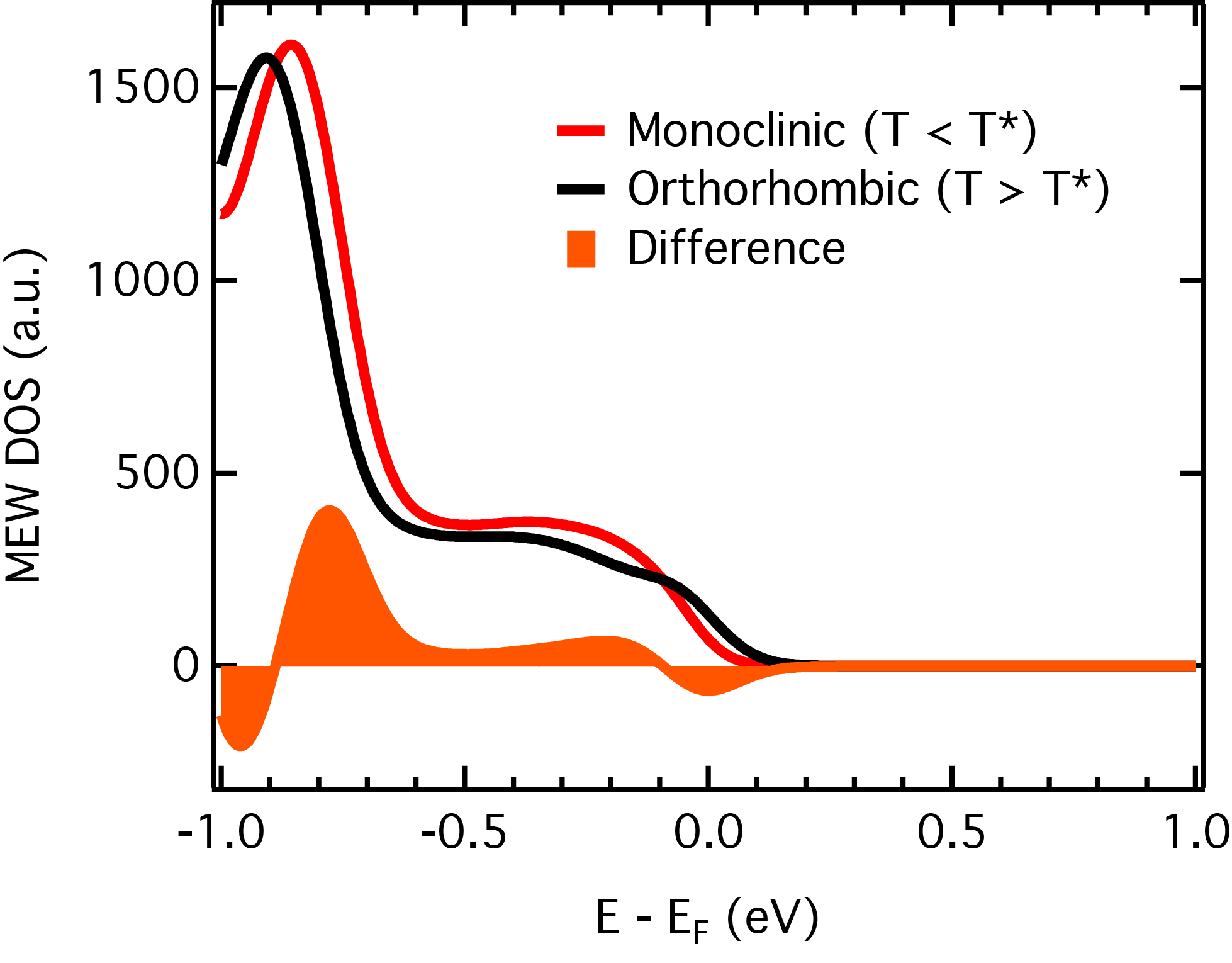}
 \caption{Matrix-element-weighted density of states obtained by taking the orbital-projected DOS from DFT, weighting them by the corresponding photoelectric cross sections for each orbital, and smoothing the result by the total experimental resolution. 
 \label{Fig_MEW_XPS}}
 \end{figure}
 
 \setcounter{table}{0}
  \setcounter{figure}{1}
  
\begingroup
\squeezetable
\begin{table}[H]
\begin{ruledtabular}
\centering
\caption{\label{tab:orth_lat_par} Calculated crystallographic parameters for orthorhombic ($Pbnm$) and monoclinic ($P2_1/n$) CaFeO$_3$ obtained from the fully optimized structures within the PBEsol exchange-correlation functional with the Dudarev form of the Hubbard correction (U = 3 eV).}
\begin{tabular}{lcccc}%
\multicolumn{5}{l}{\textbf{Orthorhombic} \hfill  $a=5.294, b=5.321, c=7.495$~\AA}  \\
\multicolumn{5}{l}{$Pbnm$ (62) \hfill $\alpha=\beta=\gamma=90^\circ$}\\[0.4ex]
Atom    &    Wyck.\ Site    & $x$    & $y$    & $z$     \\
\hline
Ca    &4c    &0.00775    & -0.0407    &$\frac{1}{4}$ \\
Fe    &4b    &$\frac{1}{2}$    &0        &0         \\
O(1)    &8d    &-0.288    &0.288    &-0.463 \\
O(2)    &4c    &-0.0711    &-0.487    &$\frac{1}{4}$ \\[0.2em]
\hline\hline\\[-1em]
\multicolumn{5}{l}{\textbf{Monoclinic} \hfill  $a=12.958, b=5.327, c=5.291$~\AA}  \\
\multicolumn{5}{l}{$P2_1/n$ (14) \hfill $\alpha=\gamma=90^\circ, \beta=144.68^\circ$}\\[0.4ex]
Atom    &    Wyck.\ Site    & $x$    & $y$    & $z$     \\
\hline
Ca    &4e    &-$\frac{1}{4}$    &0.0414    &0.492 \\
Fe(1)    &2b    &0        &0        &$\frac{1}{2}$     \\
Fe(2)    &2c    &$\frac{1}{2}$    &0        &$\frac{1}{2}$     \\
O(1)    &4e    &0.0377    &-0.285    &0.367 \\
O(2)    &4e    &0.0375    &0.209    &0.290 \\
O(3)    &4e    &-0.253    &0.486    &-0.435 \\
\end{tabular}
\end{ruledtabular}
\end{table}
\endgroup

\end{document}